\documentclass{nature}

\usepackage{lineno}
\usepackage{lmodern}
\usepackage{amsmath}
\usepackage{amssymb}
\usepackage{graphicx}
\usepackage{hyperref}
\usepackage{xspace}
\usepackage{caption}
\usepackage{mathptmx}
\usepackage{mathrsfs}
\usepackage{pdfpages}
\usepackage{mathtools} 
\usepackage{bm}
\usepackage{upgreek}

\usepackage{xr-hyper}
\makeatletter
\newcommand*{\addFileDependency}[1]{
  \typeout{(#1)}
  \@addtofilelist{#1}
  \IfFileExists{#1}{}{\typeout{No file #1.}}
}
\makeatother

\newcommand*{\myexternaldocument}[1]{
    \externaldocument{#1}
    \addFileDependency{#1.tex}
    \addFileDependency{#1.aux}
}

\myexternaldocument{supp_info}

\newcommand{\Tx}{\ensuremath{T_x}\xspace}

\title{Breaking through the Mermin-Wagner limit in 2D van der Waals magnets} 
\author{
Sarah~Jenkins$^{1,2,3}$, Levente~R\'ozsa$^{4}$, Unai Atxitia$^5$, 
Richard~F.~L.~Evans$^{1}$, Kostya S. Novoselov$^{6}$, 
Elton J. G. Santos$^{7,8\dagger}$
}

\makeatletter
\let\saved@includegraphics\includegraphics
\AtBeginDocument{\let\includegraphics\saved@includegraphics}

\makeatother

\begin{document}

\maketitle

\begin{affiliations}

\item Department of Physics, University of York, York, YO10 5DD, UK
\item TWIST Group, Institut f\"ur Physik, Johannes Gutenberg Universit\"at, 55128 Mainz, Germany
\item TWIST Group, Institut f\"ur Physik, Universit\"at Duisburg-Essen, Campus Duisburg, 47057 Duisburg, Germany
\item Fachbereich Physik, Universit\"{a}t Konstanz, D-78457 Konstanz, Germany
\item Dahlem Center for Complex Quantum Systems and Fachbereich Physik,  Freie Universit\"{a}t Berlin,  14195 Berlin, Germany 
\item Institute for Functional Intelligent Materials, National University of Singapore, 117544, Singapore
\item Institute for Condensed Matter Physics and Complex Systems, School of Physics and Astronomy, The University of Edinburgh, EH9 3FD, United Kingdom
\item Higgs Centre for Theoretical Physics, The University of Edinburgh,  EH9 3FD,  United Kingdom \\ 
$^{\dagger}$Corresponding author: esantos@ed.ac.uk  
\end{affiliations}


\date{}

\begin{abstract}
{
The Mermin-Wagner theorem~\cite{Mermin1966} states that long-range {magnetic} order does not exist in one- or two-dimensional (2D) isotropic magnets with short-ranged interactions.  
The theorem has been a milestone in magnetism and has been driving the research of recently discovered 2D van der Waals (vdW) magnetic materials\cite{gong_discovery_2017,firstCrI3} from fundamentals up to potential applications\cite{Genome22}.   
In such systems, the existence of magnetic ordering is typically attributed to the presence of a significant magnetic anisotropy,  which is known to introduce a spin-wave gap and circumvent the core assumption of the theorem\cite{Bogolyubov,Mermin1966,Hohenberg}. Here we show that 
in finite-size 2D vdW magnets typically found in lab setups ({\it e.g.,} within millimetres), 
short-range interactions can be large 
enough to allow the stabilisation of magnetic order 
at finite temperatures without any magnetic anisotropy for practical 
implementations. We demonstrate that magnetic 
ordering can be created in flakes of 2D materials independent of the lattice symmetry 
due to the intrinsic nature of the spin exchange interactions and 
finite-size effects in two-dimensions. 
Surprisingly we find that the crossover temperature, where the intrinsic magnetisation 
changes from superparamagnetic to a completely disordered paramagnetic regime, is weakly 
dependent on the system length, requiring giant sizes ({\it e.g.,} of the order of 
the observable universe $\sim$10$^{26}$ m) 
in order to observe the vanishing of the magnetic order at cryogenic temperatures as expected from the Mermin-Wagner theorem.
Our findings indicate exchange interactions as the main driving force behind 
the stabilisation of short-range order in 2D magnetism and broaden 
the horizons of possibilities for exploration of compounds with 
low anisotropy at an atomically thin level. 
}
\end{abstract}
\maketitle

\section*{Introduction}
The demand for computational power is increasing exponentially, 
following the amount of data generated across different devices, applications 
and cloud platforms\cite{Roy20,Sander_2017}. To keep up with this trend, 
smaller and increasingly energy-efficient devices must be 
developed which require the study of compounds not yet 
explored in data-storage technologies. 
The discovery of magnetically stable 2D vdW materials 
could allow for the development of spintronic devices with 
unprecedented power efficiency and computing capabilities 
that would in principle address some of these challenges\cite{Genome22}. 
Indeed, the magnetic stability of vdW layers has been one of the central 
limitations for finding suitable candidates, given that 
strong thermal fluctuations are able to rule out any magnetism. 
As it was initially pointed out by Hohenberg\cite{Hohenberg} for a superfluid or a
superconductor, and extended by Mermin and Wagner\cite{Mermin1966} for spins on a lattice, 
long-range 
order should be suppressed at finite temperatures 
in the 2D regime, when only short-range isotropic 
interactions exist. 
Importantly, the theorem only excludes long-range magnetic order at finite 
temperature in the thermodynamic limit\cite{Mermin1966}, {\it i.e.,} for infinite system sizes.
However, the common understanding is that the theorem also 
excludes the alignment of spins in samples studied experimentally 
which are a few micrometers in size\cite{Wahab2020,Geim19}, 
suggesting that such systems are indistinguishable from infinite. 
Previous reports\cite{Kapikranian_2007,Palle_2021,DeJongh01,POMERANTZ1984556,Lewenstein06,Sacha08,Lewenstein10,Crawford11,Crawford_2013,Imry75} have discussed at different levels of theoretical and experimental approaches the limitations and the potentials ways to overcome the Mermin-Wagner theorem which provide a historical evolution of the common concepts used in the field of 2D magnetism.   

The long-range order characterising infinite systems 
only becomes distinguishable from short-range order describing 
the local alignment of the spins if the system size exceeds the 
correlation length at a given temperature\cite{Halperin19}. 
Previous numerical studies and the scaling analysis of 
2D Heisenberg magnets\cite{Stanley1966,Shenker1980,Blote2002,Tomita2014} 
have established that although only short-range order is observable 
at finite temperature, the spin correlation length can be larger than the system size 
below some finite crossover temperature. 
An intriguing question on this long-range limit is how can we understand real-life materials, 
which routinely have a finite size $L$ (Fig. \ref{fig:honeycomb}{\bf a}), in light of the Mermin-Wagner theorem. It is known that thermal fluctuations will affect the emergence of a spontaneous magnetisation at low dimensionality. Nevertheless, it is unclear which kind of spin ordering can be foreseen in thin vdW layered compounds when finite-size effects and exchange interactions play together. With recent advances in computational 
power and parallelisation scalability, it is  possible to 
directly model magnetic ordering processes and dynamics of 2D materials 
on the micrometre length-scale accessible experimentally. 
Here, we demonstrate that short-range order can exist in 
systems with no anisotropy even down to the 2D limit 
using computer-intensive atomistic simulations and analytical models. 

\section*{Results}

We start by defining the magnetization in our systems as: 
\begin{equation}
    \mathbf{m}=\frac{1}{N}\sum_{i}\mathbf{S}_{i},
\end{equation}

\noindent where $\mathbf{S}_{i}$ denotes the classical spin unit vector 
at lattice site $i$ and $N$ is the number of sites. In the absence of 
external magnetic fields, the expectation value of the 
magnetization $\langle\mathbf{m}\rangle$ vanishes in any 
finite-size system due to time-reversal invariance. 
Yet, 3D systems of only a few nanometres in size that are far from infinite have been studied for decades 
and exhibit a clear crossover from a magnetically ordered to a paramagnetic phase \cite{Roduner06,Singh13}. 
The Mermin-Wagner theorem establishes that $\langle\mathbf{m}\rangle$ must also be zero in 
infinite 2D systems with short-ranged isotropic interactions. 
However, for practical implementations it is relevant to unveil
whether the average magnetisation vanishes because the spins 
are completely disordered at any point in time, or if they 
are still aligned on short distances but the overall 
direction of the magnetisation $\mathbf{m}$ strongly 
suffers time-dependent variation. 
Short-range order may be characterised by 
the intrinsic magnetisation\cite{Kachkachi2001}:
 
\begin{equation}
\langle\left|\mathbf{m}\right|\rangle=\left\langle\sqrt{\left(\frac{1}{N}\sum_{i}\mathbf{S}_{i}\right)^{2}}\right\rangle,
\end{equation}
which is always positive by definition. The intrinsic magnetisation 
is $\langle |\mathbf{m}| \rangle \approx 1$ in the short-range-ordered regime 
and converges to zero when the spins become completely disordered.\cite{Joerg21,Wahab21,Wahab2020}

\begin{figure*}[ht]\centering
\includegraphics[width=1\columnwidth]{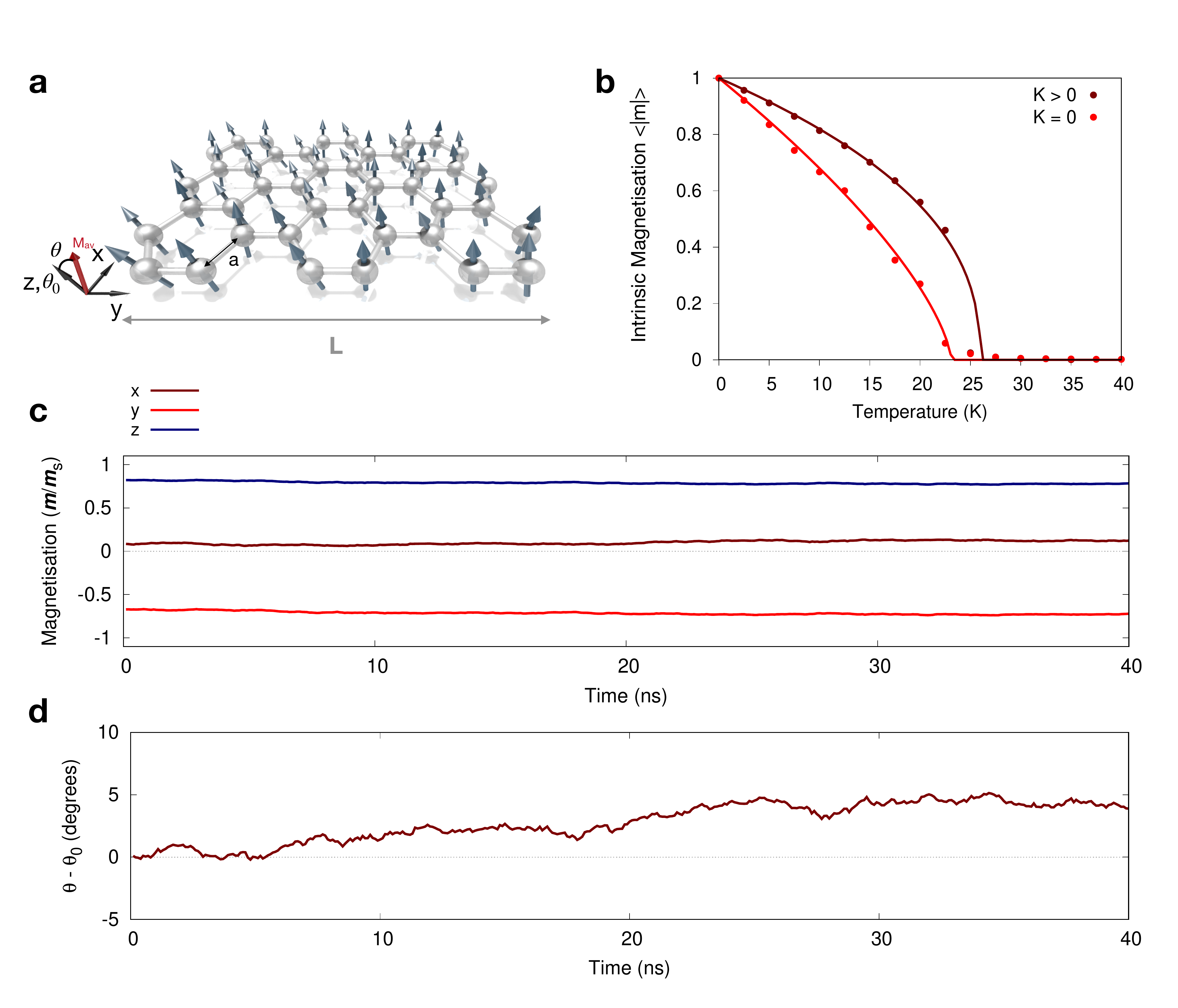}
\caption{\textbf{Short-range magnetic ordering at finite temperatures in a 2D isotropic magnet.} 
{\bf a,} Local view of the spin directions extracted from the atomistic simulations on a 2D honeycomb lattice. $a$ is the atomic spacing ($a = 0.4$ nm), L is the length considered in the computations, and \textbf{M}$_{av}$ is the averaged magnetisation vector. $\theta$ corresponds to the angle between \textbf{M}$_{av}$ and the $z$-axis. $\theta_0=0$ denotes the initial configuration aligned with the $z$-axis. 
%
{\bf b,} Temperature-dependent intrinsic magnetisation ($\langle |\mathbf{m}| \rangle$) with ($K = 1 \times 10^{-24}$ J/atom) and without ($K=0$) anisotropy in a ${1000 \times 1000}$ nm$^{2}$ flake. Solid lines are the fit to Eq.~\eqref{eq:MvT}. For $K=0$, the fitting parameters are $\beta = 0.54 \pm 0.020$ and $T_x = 23.342 \pm 0.237$ K. For $K>0$, $\beta = 0.427 \pm 0.021$ and $T_x = 26.543 \pm 0.320$ K. 
{\bf c-d,} Temporal variation of the magnetisation ($m/m_s$) and 
angle $\theta - \theta_0$, respectively, at $T = 10$~K. 
All three spatial components ($x,y,z$) are considered in {\bf c}. 
The dashed line in {\bf d} shows the initial state 
in the simulations. 
%
}
\label{fig:honeycomb}
\end{figure*}

For simplicity we first consider a 2D honeycomb 
lattice (Fig.~\ref{fig:honeycomb}{\bf a}) to model 
the magnetic ordering process for a large 
flake of $1000 \times 1000$ nm$^2$. Such a symmetry is very common in 
several vdW materials holding magnetic properties and interfaces\cite{Genome22,Zhimei22}, 
such as Cr$_2$Ge$_2$Te$_6$ (CGT) or CrI$_3$   
in which 2D magnetic ordering was first discovered\cite{gong_discovery_2017,firstCrI3}.
The system consists of 8 million atoms with nearest-neighbor Heisenberg 
exchange interactions $J_{ij}$ and no magnetic anisotropy ($K$) 
using highly accurate Monte Carlo simulations 
(see Supplementary Sections 1-2 for details). 
We use an isotropic Heisenberg spin Hamiltonian 
$ \mathcal{H} = -\sum_{i<j} J_{ij} \mathbf{S}_i \cdot \mathbf{S}_j $ 
as stated in the Mermin-Wagner theorem\cite{Mermin1966}. 
As it is shown below,  our conclusions do not depend on 
the magnitude of the exchange interactions chosen. Nevertheless, 
to give a flavor of a potential material to study, we set $J_{ij}$ to similar values to those obtained for 
CGT layers\cite{gong_discovery_2017} where a negligible magnetic anisotropy ($<1~\mu$eV) 
was observed for thin layers but yet a stable magnetic signal was 
measured at finite temperatures ($\sim$4.7 K).  
We begin by assessing the existence of any magnetic order at 
non-zero temperatures by equilibrating the system 
for $39 \times 10^{6}$ Monte Carlo steps using a 
uniform sampling\cite{Alzate-CardonaJPCM2019} 
to avoid any potential bias before 
a final averaging at thermal equilibrium for a 
further $10^6$ Monte Carlo steps. 


Strikingly, a crossover between the low-temperature short-range-ordered 
regime and the completely disordered state ($\langle |\mathbf{m}| \rangle \approx 0$) 
is observed at nonzero temperatures (Fig.~\ref{fig:honeycomb}{\bf b}) and zero magnetic anisotropy ($K=0$). 
To estimate the crossover temperature ($T_x$), 
the simulation data was fitted by the Curie-Bloch equation in the classical limit\cite{Wahab2020}:
\begin{equation}
    \langle|\mathbf{m}|\rangle(T) =  \left( 1- \frac{T}{T_x} \right)^\beta,
    \label{eq:MvT}
\end{equation}
where $T$ is the temperature and $\beta$ is an   
exponent in the fitting. From the fitting one obtains 
$T_x = 23.342 \pm 0.237$ K ($\beta = 0.54 \pm 0.020$), 
which is about one third of the mean-field (MF) critical temperature $T_{\textrm{c}}^{\rm{MF}}=zJ_{ij}/\left(3k_{\textrm{B}}\right)=70.8$~K (where $z = 3$ 
is the number of nearest neighbours) even for this considerable system size. 
The simulations were then repeated including magnetic anisotropy 
($K = 1 \times 10^{-24}$ J/atom) which resulted in a slight increase in 
the crossover temperature ($T_x = 26.543 \pm 0.320$ K, $\beta = 0.427 \pm 0.021$) (Fig.~\ref{fig:honeycomb}{\bf b}). 
We observed that this difference in $T_x$ between isotropic and anisotropic 
cases becomes negligible as the flake size is reduced (100 $\times$ 100 nm$^2$) 
with minor variations of the curvature of the magnetisation versus 
temperature (Supplementary Section 3 and Supplementary Figure 1). 
We also checked that different Monte Carlo sampling 
algorithms ({\it i.e.,} adaptive) and starting spin 
configurations ({\it i.e.,} ordered, disordered) do not 
modify the overall conclusions (Supplementary Section 4 and Supplementary Figure 2). 
Taking dipolar interactions into account only has a minor effect on the intrinsic magnetization curve (Supplementary Figure 3).  
Although the magnetocrystalline anisotropy $K$ or the dipolar interactions circumvent the Mermin-Wagner theorem 
and lead to a finite critical temperature, this indicates that 
systems up to lateral sizes of 1 $\mu$m are not suitable for 
observing the critical behaviour. Instead the crossover in the 
short-range order defined by the isotropic interactions 
dominates in this regime, regardless of whether the anisotropy is present or absent. 
Previous studies on finite magnetic clusters on metallic 
surfaces\cite{Ebert06,Ebert07} suggested that anisotropy is not the key factor 
in the stabilisation of magnetic properties at low dimensionality and finite temperatures,  
but rather it determines the orientation of the magnetisation. 

Even though short-range interactions can stabilise short-range 
magnetic order in 2D vdW magnetic materials, this does not necessarily 
imply that the direction or the magnitude of the magnetisation 
is stable over time. As thermally activated magnetisation 
dynamics may potentially change spin directions\cite{Fuller63}, 
it is important to clarify whether angular variations of the spins are present. 
Hence we compute the time evolution of the magnetisation 
along different directions ($x,y,z$) and its angular dependence (Fig. \ref{fig:honeycomb}{\bf c,d}) 
through the numerical solution of the Landau-Lifshitz-Gilbert equation (see Methods for details). 
Over the whole simulation (40 ns), all components of the 
magnetisation assume approximately constant  
values which deviate by $\pm 5^\circ$ from the 
mean direction $\theta_{\mathrm{av}}$. Similar 
analyses undertaken for different flake sizes 
($L \times L$, $L=$ 50, 100, 500 nm) show that the 
spin direction is very stable at each temperature 
considered (2.5 K, 10 K, 20 K, 30 K, 40 K) and 
follows a Boltzmann distribution (Supplementary Section 5 and Supplementary Figure 4). 
These results show that the magnetisation in a 
2D isotropic magnet is not only stable in magnitude 
but its direction only negligibly varies over time. 

\begin{figure*}[h]\centering
\includegraphics[width=1\columnwidth]{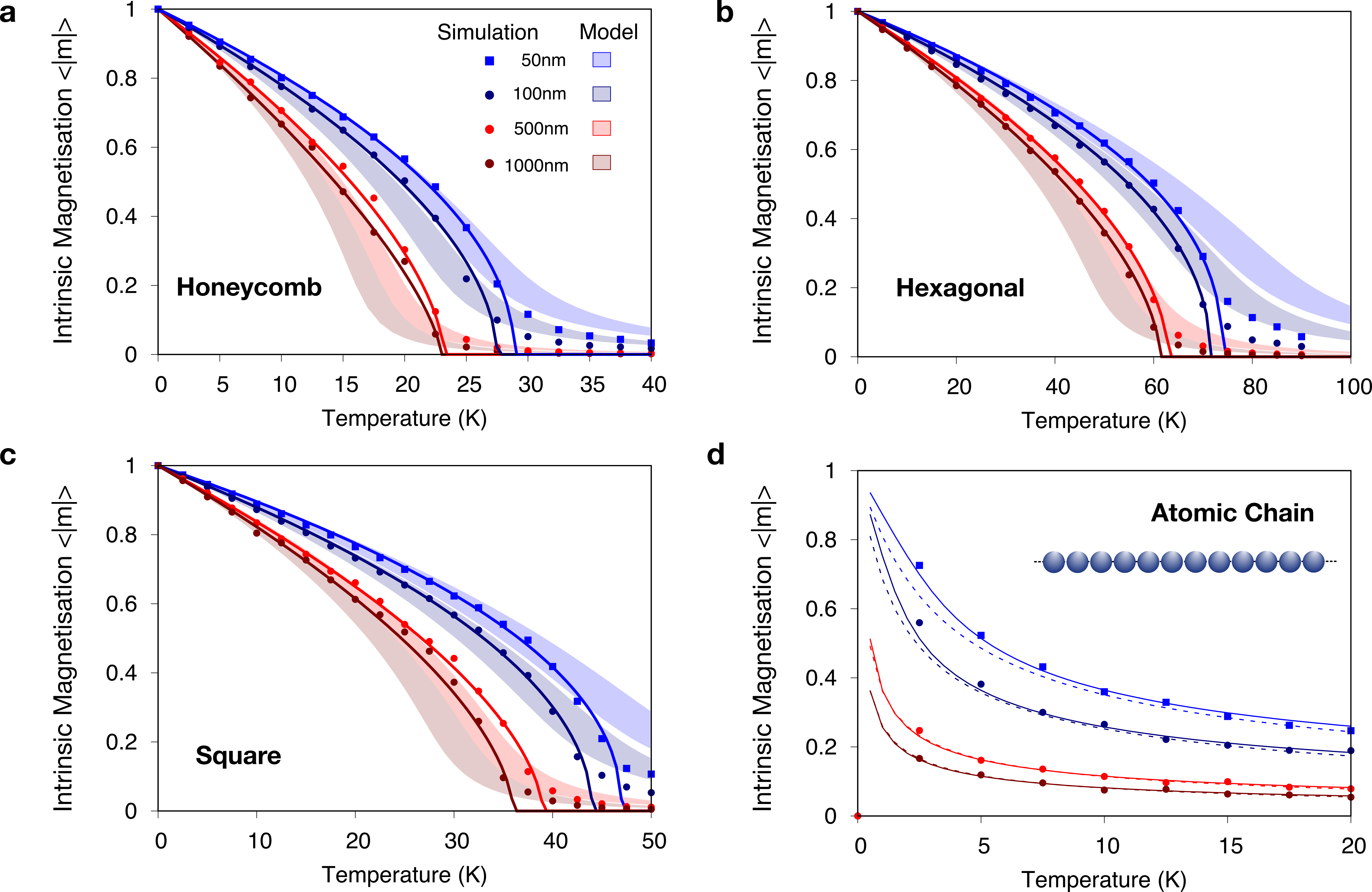}
\caption{\textbf{Temperature- and size-dependent properties of isotropic 1D and 2D materials with different crystal structures.} 
{\bf a-d} Comparative simulations of the temperature-dependent magnetisation for honeycomb, hexagonal, square lattices and an atomic chain (1D), respectively, for different system sizes. Points indicate the results of Monte Carlo simulations, the lines show fits to the Curie-Bloch Eq. \eqref{eq:MvT} in the classical limit, and the shaded regions indicate the 
anisotropic spherical model calculations for different assumptions of the renormalisation factor for the Curie temperature arising from the mean-field approximation. See Supplementary Section 7 for details.  
The dashed and solid lines in {\bf d} 
indicate the anisotropic spherical model calculations, 
and the exact solution, respectively. 
Both show a sound agreement with the atomistic simulations. 
The datasets in {\bf a-c} clearly show the existence of short-range collinear magnetic order for all 2D lattices at the simulated sizes considered with nonzero crossover temperature. Zero magnetic anisotropy is included in all calculations. 
}
\label{fig:MvT}
\end{figure*}

An outstanding question raised by the modelling of the 2D finite 
flakes is whether other kind of common lattice symmetries 
({\it i.e.,} hexagonal, square), lower dimensions ({\it i.e.,} 1D) 
and different sizes may follow similar behaviour to 
that found in the honeycomb lattice. Figure \ref{fig:MvT} 
shows that the effect is universal regardless of 
the details of the lattice or the dimension considered. 
%
We find persistent magnetic order for $T > 0$~K at zero magnetic anisotropy 
for the cases considered. There is a consistent reduction in the crossover temperature as a 
function of the system size $L \rightarrow \infty$ in 
agreement with the general trend of the temperature dependence 
of the correlation length discussed above (Fig. \ref{fig:MvT}{\bf a-c}). 
The 1D model (atomic chain) displays a similar 
trend (Fig. \ref{fig:MvT}{\bf d}) although the variation 
of $\langle|\mathbf{m}|\rangle$ with $T$ is different 
due to the lower dimensionality. 
We have also checked that several additional factors do not affect these conclusions, such as i) the type of boundary conditions, {\it e.g.,} open; ii) flake shape ({\it e.g.,} circular), and iii) strength of the exchange interactions. Supplementary Figures 5-6 provide a summary of this analysis. Indeed, the stabilisation of magnetism in 2D is independent of the magnitude of the exchange interactions considered, as a linear re-scaling of the temperatures is obtained for different $J_{ij}$ values. This indicates the generality of the results which are valid regardless of the chemical details of the 2D material and its corresponding $J_{ij}$ interactions. Moreover, if the exchange coupling between atoms could be engineered via chemical synthesis\cite{Klimov09,Long11,Baumgarten19}, then magnets with either low or high crossover temperatures might be fabricated depending on the target application. Such procedure would not require heavy elements with sizeable spin orbit-coupling for the generation of magnetic anisotropy since it is not necessary for 2D magnetism.


%

%
To give an analytical description  of these effects, we use the 
anisotropic spherical model (ASM) for the calculation of 
the finite-size effects on the intrinsic 
magnetization\cite{Garanin1996,Garanin1999,Kachkachi2001} 
(see Supplementary 
Section 7 for details). 
The ASM takes into account Goldstone modes in the system 
and self-consistently generates a gap in the 
correlation functions which avoids infrared divergences 
responsible for the absence of long-range order for isotropic 
systems in dimensions $d \leq 2$ as $L \rightarrow \infty$ 
as per the Mermin-Wagner theorem. 
We applied the formalism to 1D and 2D systems for the isotropic 
Heisenberg Hamiltonian in the absence of an external 
magnetic field\cite{Kachkachi2001}. The results of 
our analytical calculations are shown as shaded regions 
in Fig.~\ref{fig:MvT} (see Supplementary Section 6 for the definition of the regions). 
At low temperatures both limits agree well with our 
Monte Carlo calculations within the statistical noise 
and clearly show the existence of a finite intrinsic magnetisation 
at non-zero temperature for finite size. At higher temperatures 
there is a systematic difference between the degree of magnetic 
ordering between the simulations and the analytical calculations 
due to the ASM only becoming exact in the limit of 
infinitely many spin components. 
The large number of Monte Carlo steps and strict convergence criteria 
to the same thermodynamic equilibrium for ordered and disordered 
starting states (Supplementary Section 4) rule 
out critical slowing down\cite{NightingalePRL1996} as a 
source of difference between the analytical calculations and the simulations. 

\begin{figure*}[h]\centering
\includegraphics[width=0.6\columnwidth]{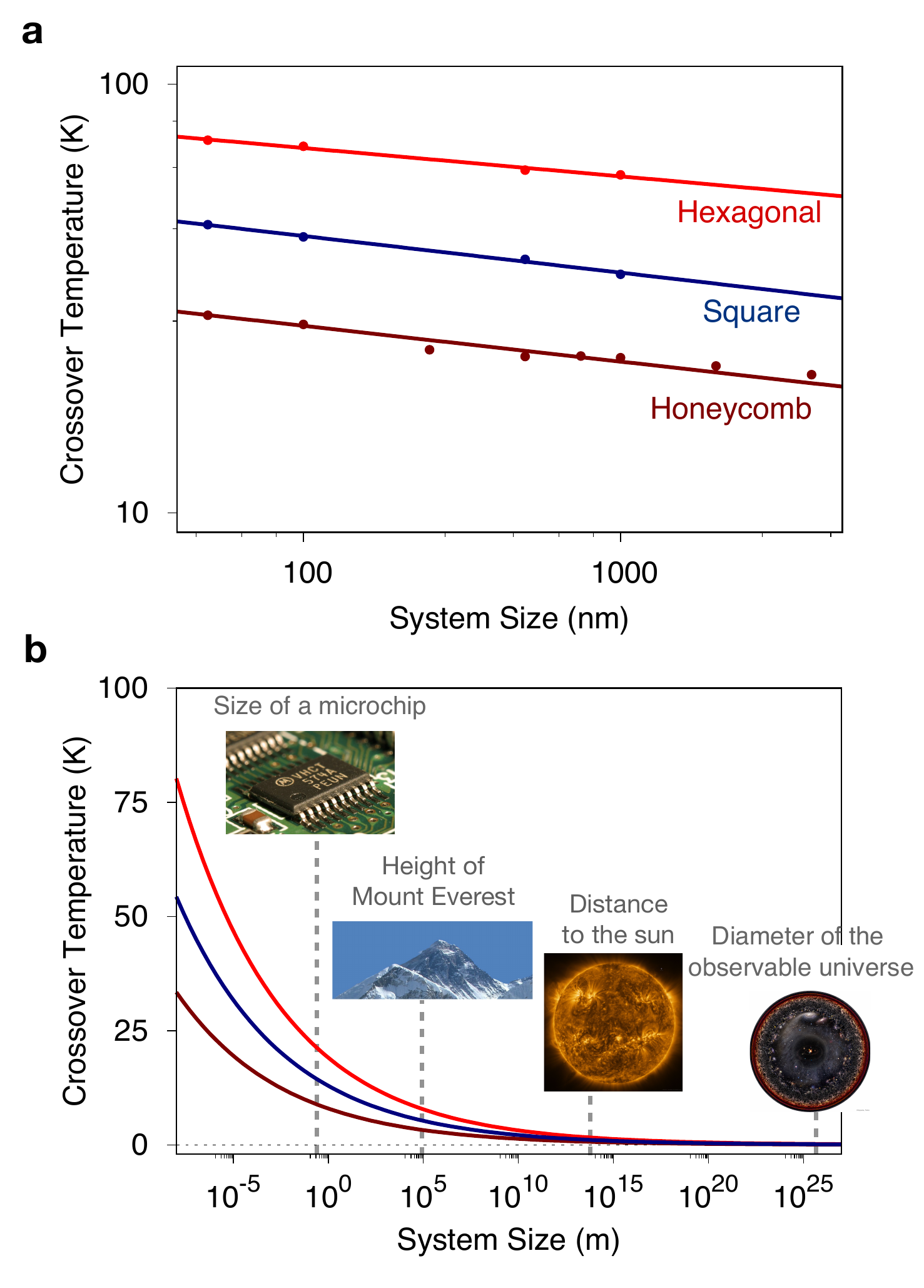}
\caption{\textbf{Size scaling of the simulated crossover temperature for the different 2D lattices}
{\bf a,} Variation of the crossover temperature T$_x$ with system size for different symmetries (Hexagonal, Square, Honeycomb) on a log-scale. The curves are a fit using $T_x=A/\log(L/B)$, where $A$ and $B$ are fitting constants and $L$ is the system size. $A$ and $B$ are 327.28 K and 0.000542 nm, 484.96 K and 0.00166 nm and 1018.50 K and 5.7$\times$10$^{-5}$ nm for honeycomb, square and hexagonal lattices, respectively. 
{\bf b,} Extrapolation of the exponential fits in {\bf a} to larger sizes on all studied symmetries. The crossover temperature remains finite ($>$4 K) for systems as large as  $\sim$10$^{25}$ m indicating no dependence of the magnetic anisotropy for stabilisation of magnetic ordering. Insets provide comparison with physical distances observed in different systems. 
Figures in {\bf b} are adapted with permission under a Creative Commons CC BY license from Wiki Commons. Microchip: Integrated circuit on microchip by Jon Sullivan, 2006, at Public Domain from Wiki Commons. 
Sun: inset is from ESA \& NASA/Solar Orbiter/EUI team, 2022 at Public Domain from Wiki Commons. Data processing by E. Kraaikamp. 
Everest: Wikivoyage banner for Mount Everest or Nepal by Fabien1309. This file is made available under the Creative Commons CC0 1.0 Universal Public Domain Dedication. 
Universe: The Observable Universe by Pablo Carlos Budassi from Wikipedia under Attribution-ShareAlike 3.0 Unported (CC BY-SA 3.0). 
}
\label{fig:universe}
\end{figure*}

One may also argue in terms of the correlation length $\xi$ which is comparable to the system size at the crossover temperature. 
It has been demonstrated\cite{Shenker1980} that $\xi\propto\textrm{exp}(cJ/T)$, where $c$ is a constant, meaning that the inverse crossover temperature $T_x^{-1}$ only logarithmically increases with the system size. Although our simulations are at the limit of the capabilities of current supercomputer features, this effect is expected to persist for larger sizes of 2$-$10 $\upmu$m. These values represent typical sizes of continuous 2D microflakes in experiments, and much larger than the ideal nanoscale devices likely to be used in future 2D spintronic applications. Fitting a scaling function to the crossover temperatures for different lattice symmetries (Fig.~\ref{fig:MvT}), we can plot the scaling of the crossover temperature with size (Fig. \ref{fig:universe}{\bf a}) which can then be extrapolated to larger scales. The crossover temperature is still approximately 30 K for $2 - 10$ $\upmu$m flakes (Fig.\ref{fig:universe}{\bf b}).
The graph can be extrapolated to show that only at the 10$^{15}-10^{25}$ m range does the crossover temperature become lower than $\sim$1 K. To put these numbers into perspective for physical systems, these length scales lie between the distance of the Earth to the Sun and the diameter of the observable universe. Therefore, the often asserted notion\cite{Genome22} that experimental 2D magnetic samples can be classified as infinite and therefore display no net magnetic order at nonzero temperatures, as expected from the Mermin-Wagner theorem, is not applicable. Surprisingly, simple estimations by Leggett\cite{Leggett13} for the stability of graphene crystals following the Mermin-Wagner theorem would require sample sizes of the order of the distance from the Earth to the Moon which are in sound agreement with our simulation results.  

\begin{figure*}[!tb]
\includegraphics[width=15.8cm]{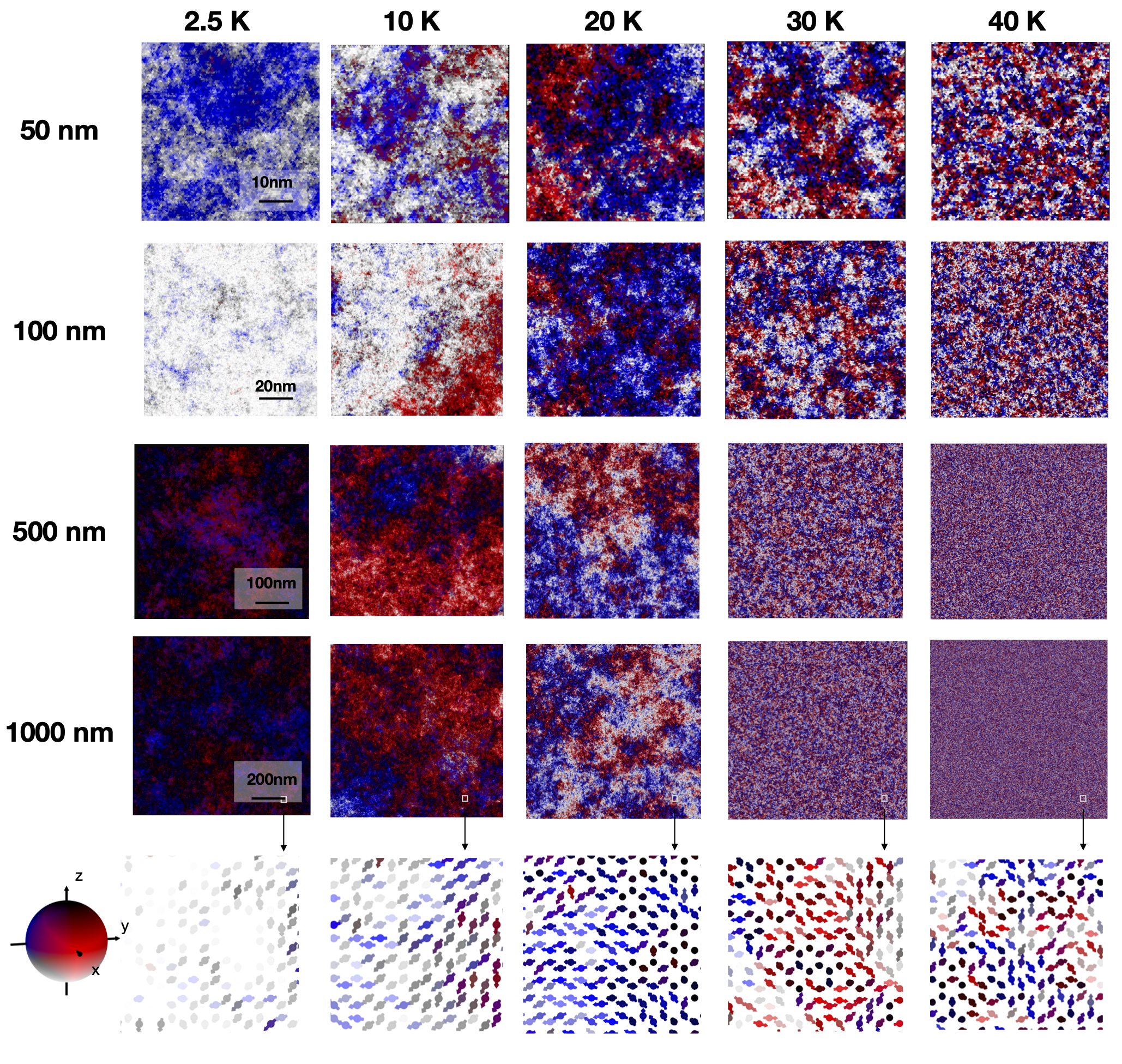}
\caption{\textbf{Temperature-dependent magnetic order.} Visualisations of the magnetic spin configurations for the honeycomb lattice starting from an ordered state as a function of system size (vertical row) and temperature (horizontal row). The spins are projected following the color scale shown in the sphere on the left. The bottom row shows a local view of the spins inside a 5 nm $\times$ 5 nm area at the location outlined by the small boxes in the $1000 \times 1000$ nm$^2$ snapshots.}
\label{fig:crystal}
\end{figure*}

The significance of the crossover temperature $\Tx$ in relation 
to the Curie temperature $T_{\mathrm{C}}$ is particularly important 
when discussing the nature of the 
magnetic ordering in 2D magnets 
at zero anisotropy for $T>0$ K. 
We investigate this behaviour through colour 
maps of the spin ordering after 40 million Monte Carlo steps 
comparing different system sizes and temperatures (Fig.~\ref{fig:crystal}). 
At very low temperatures $T = 2.5$~K, where there is a high degree of order, 
the spin directions are highly correlated, as indicated by a mostly uniform colouring. 
Although the temperatures are near zero, the system is superparamagnetic 
indicating that over time the magnetization direction fluctuates, 
and the effect is most apparent for the smallest sizes where 
the average direction has moved significantly from the initial 
direction $\mathbf{S} || z$. At higher temperatures the deviation of the 
spin directions within the sample increases as indicated by 
the more varied colouring. To quantitatively assess the spin 
deviations we plot the statistical distribution of angle between 
the spin direction and the mean direction for different 
temperatures for each size (Supplementary Figure 4). 
For an isotropic distribution on the unit sphere there 
is a $\sin(\theta)$ weighting which is seen at the highest 
temperature for all system sizes. For lower temperatures 
where the spin directions are more correlated, the distribution 
is biased towards lower angles. Qualitatively there is little difference 
in the spin distributions for the different samples. At $T = 20$~K there is 
however a systematic trend in the peak angle increasing 
from $\theta = 40^\circ$ for the $50 \times 50$~nm$^2$ flake 
(Supplementary Figure 4{\bf a}) to around $\theta = 60^\circ$ at
$1000 \times 1000$~nm$^2$ (Supplementary Figure 4{\bf d})
indicating an increased level of disorder 
averaged over the whole sample. 
This effect is straightforwardly explained by 
the size dependence of spin-spin correlations (Supplementary Figure 7). 
At small sizes the spins are strongly exchange 
coupled, preventing large local deviations of the 
spin directions. At longer length scales available 
for the larger systems, the variations in the magnetisation 
direction are also larger. Surprisingly, our calculations 
reveal that this effect is weak: even for very large 
flakes of a micrometre in size, only a small increase 
can be observed in the position of the peak in the angle 
distribution at a fixed temperature. Above the crossover 
temperature 
the spin-spin correlation length 
becomes very small compared to the system size with rapid local 
changes in the magnetisation direction, indicative of a 
completely disordered paramagnetic state. Our analysis 
reveals that the spins in finite-sized 2D isotropic 
magnets are strongly aligned due to short-range order 
at non-zero temperatures and up to the crossover temperature.

\section*{Discussion}

Mathematically a phase transition is defined as a {non-analytic} 
change in the state variable for the system, such as the 
particle density or the magnetization in the case of spin systems. 
For any {finite} system the state variable is continuous by 
definition due to a finite number of particles, forming a continuous 
path of intermediate states between two distinct physical phases\cite{Stanley1971}. 
The same is true for a magnetic system, forming a continuous path between an ordered and a paramagnetic state.  
A priori then, it is impossible to have a true phase transition 
for any finite magnetic samples which are 
routinely implemented in device platforms.  
Yet, nanoscale magnets that are far from 
infinite have been studied for decades and exhibit 
a clear crossover from magnetically ordered to paramagnetic 
phases, occurring for systems only a few nanometres in size\cite{Roduner06,Singh13}. 
The crossover temperature in a finite-size system hence can be described as an 
inflection point in M(T). 
The precise definition of a phase transition  is 
significant when considering the main conclusions 
of Mermin and Wagner\cite{Mermin1966}, which explicitly 
only apply in the case of an infinite system. 
As our results clearly show, 
sample sizes measured experimentally are not classifiable 
as infinite and therefore not subject to the Mermin-Wagner theorem. 
It is noteworthy that 3D compounds have weak dependence of their critical temperature on the magnetic anisotropy\cite{Coey2010}. Similar analysis performed for a finite 3D bulk system (Supplementary Figure 8{\bf a-b}) show that the inclusion of anisotropy barely change the results for T$_{\rm c}$. This suggests that magnetism is an exchange-driven effect in both two and three dimensions.



On the practical side, heterostructures with conventional 
metallic magnetic materials could establish preferential 
directions of the magnetization through anisotropic exchange 
and dipolar couplings. However, it is important to point out that 
the short-range order is enforced by the isotropic exchange couplings 
and even a low anisotropy may suffice for stabilizing the direction 
of the magnetization in the vdW layers, {\it i.e.,} from underlying 
magnetic substrates. We can imagine micrometre-sized samples 
where all spins are still correlated at finite temperatures 
so it could represent a single bit. However, 
for miniaturization purposes multiple nanometre-sized bits are required on 
the same sample in order to be implemented 
in recording media. This is typically achieved 
by magnetic domains, but there are no domains 
in an isotropic model since the domain wall width is infinite.
However, if vdW layers can be grown with grain boundaries, like 
in 2D mosaics\cite{mosaic20}, which are large enough 
that each grain area would have an uniform magnetisation, 
then a magnetic monolayer would have as many bits as 
available on the material surface. The underlying substrate hence would 
set the magnetisation direction for further implementations. 
This spin-interface engineering would be 
a considerable step towards on-demand magnetic properties at 
the atomic level given the flexibility on the orientation of the 
magnetic moments without a predefined direction at the layer. 
While the anisotropy circumvents the Mermin-Wagner 
theorem and causes the critical temperature $T_{\textrm{c}}$ to 
be nonzero in infinitely large systems, in finite samples the 
short-range order persists up to much higher temperatures  
($T_{\textrm{x}} > T_{\textrm{c}}$) since $T_{\textrm{x}}$ 
is proportional to the isotropic exchange rather 
than the anisotropy\cite{PhysRevB.60.1082,PhysRevB.71.024427}. 
Indeed, the long tail features observed in the intrinsic 
magnetisation (Fig. \ref{fig:MvT}) extending
above the crossover temperature suggest that short-range 
order is present.  
In addition, the existence of short-range order in bulk magnetic systems near and  
above the Curie temperature has been experimentally and theoretically 
discovered in elemental transition metals\cite{Feder85,Jelitto82,Antropov05}. 
These studies indicate the persistence of magnetic ordering within the 
supposedly disordered phase above the Curie temperature, where any ordered phase 
is primarily controlled by exchange interactions as in the case for 2D magnets.  
%
For instance, in bcc-Fe a short-range order within 5.4~\AA~was 
found\cite{Feder85} which is much smaller than the magnitudes obtained 
in our simulations for vdW materials.






In conclusion, we presented large-scale spin dynamics simulations 
and analytical calculations on the temperature dependence 
of the intrinsic magnetization in 2D magnetic materials described 
by an isotropic Heisenberg model. We found that short-range magnetic order 
at non-zero temperature is a robust feature of isotropic 2D magnets even 
at experimentally accessible length and time scales. Our data show that 
the often asserted Mermin-Wagner limit\cite{Mermin1966} does not 
apply to 2D materials on real laboratory sample sizes . 
Since the spins are aligned due to the exchange interactions 
already in the isotropic model, the direction of the magnetization may 
be stabilized by geometrical factors or finite-size effects. 
These findings open up possibilities for a wider range of 2D magnetic 
materials in device applications than previously envisioned. 
Furthermore, the limited applicability of the analytical Mermin-Wagner theorem opens 
similar possibilities in other fields such as superconductivity\cite{Palle_2021} 
and liquid crystal systems\cite{Illing1856}, where the relevant length 
scale of correlations is known to be much greater than 
that required for experimental measurements and applications. 
Our results suggest that if the magnetic anisotropy can be 
controlled to a certain degree\cite{verzhbitskiy_controlling_2020} 
until it completely vanishes, new effects of strongly correlated spins or 
more unusual disordered states may be observed.  



\section*{Methods}
We used atomistic simulations methods\cite{Wahab2020,Kartsev2020,Wahab21,Alliati2022,Strungaru22,Augustin2021} 
implemented in the VAMPIRE software\cite{Evans_2014} to compute the magnetic properties of 2D magnetic materials. The energy of our system is calculated using the spin Hamiltonian:
\begin{equation}
    \mathcal{H} = -\sum_{i<j} J_{ij} \mathbf{S}_i \cdot \mathbf{S}_j - K \sum_i (S^z_i)^2 ,
    \label{eq:Hamiltonian}
\end{equation}
where $\mathbf{S}_{i,j}$ are unit vectors 
describing the local spin directions on magnetic 
sites $i,j$, and $J_{ij}$ is the exchange constant 
between spins. An easy-axis magnetocrystalline anisotropy constant $K$ 
can be included as well, with negligible modifications of the 
results as described in the text. Simulations were run for system 
sizes of 50~nm, 100~nm, 500~nm and 1000~nm laterally 
along the $x$ and $y$ directions with periodic boundary conditions (PBCs), and 
1 atomic layer thick along the $z$ direction. Similar PBCs were used in the analytical model. However, results using open boundary conditions (OBCs) ended up in similar 
conclusions (Supplementary Figure 5). For the honeycomb lattice, 
the simulations were initialized 
in either a perfectly ordered state aligned along the $z$ direction 
or a random state corresponding to infinite temperature. 
For these simulations the final $\langle |\mathbf{m}| \rangle (T)$ 
curves were identical to each other. 
However, at low temperatures it took ten times as many steps to 
reach the final equilibrium state from the random state, so for the remaining structures 
only simulations starting from the ordered states were run. 
The systems were integrated using a Monte Carlo integrator 
using a uniform sampling algorithm\cite{Evans2014} 
to remove any bias introduced from more advanced 
algorithms\cite{Alzate-CardonaJPCM2019}. To investigate the 
temperature dependence, the simulation temperature was varied 
from 0 to 90 K in 2.5 K steps. 
$40\times10^{6}$ Monte Carlo steps were run for each temperature step. 
This was split into $39\times10^{6}$ equilibration steps 
and then $10^{6}$ time steps from which the statistics were calculated.
The Monte Carlo simulations use a pseudo-random number sequence 
generated by the Mersenne Twister algorithm\cite{Matsumoto1998MersenneTA} 
due to its high quality, avoiding correlations 
in the generated random numbers and with an exceptionally 
long period of $2^{19937}-1 \sim 10^{6000}$. The parallel implementation 
generates different random seeds on each processor to ensure 
no correlation between the generated random numbers. 

The time-dependent simulations in Figure\ref{fig:honeycomb}{\bf c,d} were performed by solving the stochastic Landau-Lifshitz-Gilbert equation:
\begin{equation}
    \frac{\partial \mathbf{S}_i}{\partial t} = -\frac{\gamma_e }{1 + \lambda^2} \left[ \mathbf{S}_i \times \mathbf{B}_{\mathrm{eff}} + \lambda \mathbf{S}_i \left(\mathbf{S}_i \times  \mathbf{B}_{\mathrm{eff}} \right) \right],
     \label{eq:LLG}
\end{equation}
which models the interaction of an atomic spin moment $\mathbf{S}_i$ with an effective magnetic field $\mathbf{B}_{\text{eff}} = -\partial \mathbf{\mathcal{H}} / \partial \mathbf{S}_i$. The effective field causes the atomic moments to precess around the field, where the frequency of precession is determined by the gyromagnetic ratio of an electron ($\gamma_e =1.76 \times 10^{11}$ rad s$^{-1}$T$^{-1}$) and $\lambda = 1$ is the damping constant. The large value of $\lambda$ was used to accelerate the relaxation dynamics in order to be computationally achievable ($\sim$72 hours). For a different damping, one has to wait longer or shorter for this to happen. Based on the system sizes used in our computations, this can vary between $\sim$5 days up to several weeks, which is not practical. However, once the system is at equilibrium, the value of the damping is not important as it is the case in our results. Moreover, a large damping would correspond to large fluctuations on the magnitude of the magnetization and its direction. Lower damping would lead to naturally slower dynamics of the magnetization. Nevertheless, we barely noticed any at the timescale included in our work (Fig. \ref{fig:honeycomb}{\bf c-d}). It is worth mentioning that no damping parameters is present in the Monte Carlo calculations which support our conclusions. The effect of temperature is taken into account using Langevin dynamics\cite{1060329}, where the thermal fluctuations are represented by a Gaussian white noise term. At each time step the instantaneous thermal field acting on each spin is given by
\begin{equation}
    \mathbf{B}^i_{\text{th}} =  \sqrt{\frac{2\lambda k_B T}{\gamma \mu_s \Delta t}}\boldsymbol{\Gamma}(t) 
\end{equation}
where $k_{\rm{B}}$ is the Boltzmann constant, $T$ is the system temperature and $\boldsymbol{\Gamma}(t)$ is a vector of standard (mean 0, variance 1) normal variables which are independent in components and in time. The thermal field is added to the effective field in order to simulate a heat bath. 
The system was integrated using a Heun numerical scheme\cite{Evans2014}.

\section*{Acknowledgments}
We thank David Mermin, Mikhail Katsnelson, and Bertrand Halperin for 
valuable discussions. 
L.R. gratefully acknowledges funding by the National Research, Development and Innovation Office of Hungary via Project No. K131938 and by the Young Scholar Fund at the University of Konstanz. U.A. gratefully acknowledges funding by the Deutsche Forschungsgemeinschaft (DFG, German Research Foundation)—Project-ID 328545488—TRR 227, Project No. A08; and grants PID2021-122980OB-C55 and RYC-2020-030605-I funded by MCIN/AEI/10.13039/501100011033 and by “ERDF A way of making Europe" and “ESF Investing in your future”.
E.J.G.S. acknowledges computational 
resources through CIRRUS Tier-2 HPC 
Service (ec131 Cirrus Project) at EPCC (http://www.cirrus.ac.uk) funded 
by the University of Edinburgh and EPSRC (EP/P020267/1); 
ARCHER UK National Supercomputing Service (http://www.archer.ac.uk) {\it via} 
Project d429. E.J.G.S. acknowledges the Spanish Ministry of 
Science's grant program ``Europa-Excelencia'' under 
grant number EUR2020-112238, the EPSRC Early Career 
Fellowship (EP/T021578/1), and the University of 
Edinburgh for funding support. For the purpose of open access, the authors have applied a Creative Commons Attribution (CC BY) licence to any Author Accepted Manuscript version arising from this submission. 

\section*{Supplementary Materials}

Supplementary Sections 1-6, Supplementary Figures 1-8, and Supplementary References. 

\section*{Data Availability}

The data that support the findings of this study are available within the paper and its Supplementary Information.

\section*{Competing interests}

The Authors declare no conflict of interests.

\subsubsection*{Author Contributions} 
E.J.G.S. conceived the idea and supervised the project. 
S.J. performed the atomistic simulations with inputs from E.J.G.S. and R.F.L.E.  
L.R. and U.A. developed the semi-analytical model and undertook the numerical simulations. E.J.G.S. wrote the paper with a draft initially prepared by S.J. 
and R.F.L.E. and also with inputs from K.S.N., U.A. 
and L.R. All authors contributed to this work, read the 
manuscript, discussed the results, and agreed on 
the included contents.

\pagebreak

\section*{References}
\bibliography{library}

\pagebreak 

\end{document}